# SK-PINN: Accelerated physics-informed deep learning by smoothing kernel gradients

Cunliang Pan[1], Chengxuan Li[1], Yu Liu[1], Yonggang Zheng[1,2], Hongfei Ye[1,*]

[1] State Key Laboratory of Structural Analysis, Optimization and CAE Software for Industrial Equipment, Department of Engineering Mechanics, School of Mechanics and Aerospace Engineering, Dalian University of Technology, Dalian 116024, PR China

[2] DUT-BSU Joint Institute, Dalian University of Technology, Dalian 116024, P. R. China

**Abstract**

The automatic differentiation (AD) in the vanilla physics-informed neural networks (PINNs) is the computational bottleneck for the high-efficiency analysis. The concept of derivative discretization in smoothed particle hydrodynamics (SPH) can provide an accelerated training method for PINNs. In this paper, smoothing kernel physics-informed neural networks (SK-PINNs) are established, which solve differential equations using smoothing kernel discretization. It is a robust framework capable of solving problems in the computational mechanics of complex domains. When the number of collocation points gradually increases, the training speed of SK-PINNs significantly surpasses that of vanilla PINNs. In cases involving large collocation point sets or higher-order problems, SK-PINN training can be up to tens of times faster than vanilla PINN. Additionally, analysis using neural tangent kernel (NTK) theory shows that the convergence rates of SK-PINNs are consistent with those of vanilla PINNs. The superior performance of SK-PINNs is demonstrated through various examples, including regular and complex domains, as well as forward and inverse problems in fluid dynamics and solid mechanics.

---

[*] Corresponding author, email: yehf@dlut.edu.cn

# 1. Introduction

Partial differential equations (PDEs) provide a concise and effective framework for addressing numerous problems in science and engineering. Unfortunately, most PDEs cannot be solved analytically. Over the past decades, various numerical methods have been developed to approximate PDE solutions. Early numerical methods primarily included grid-based methods such as finite difference methods (FDM) [1], finite element methods (FEM) [2], finite volume methods (FVM) [3], and spectral methods [4]. To reduce the grid dependency, a new generation of meshless numerical methods has been proposed including smoothed particle hydrodynamics (SPH) [5], discrete element methods (DEM) [6], element-free Galerkin methods (EFGM) [7], and reproducing kernel particle methods (RKPM) [8].

Deep neural networks (DNN) have recently led to significant progress in solving PDEs [9]. The physics-informed neural networks (PINNs) are among the most notable advancements [10]. The term "PINN" was coined by Raissi et al. [10], while the concept of PINNs was initially proposed in the early 1990s [11-13]. Various PINN models and applications have emerged with the rise of deep learning. Different architectures have been explored in the context of PINNs, ranging from fully-connected neural networks [10, 14] to convolutional neural networks (CNNs) [15, 16], recurrent neural networks (RNNs) [17, 18], generative adversarial networks (GANs) [19, 20], and recently proposed Kolmogorov-Arnold networks (KANs) [21, 22]. PINNs have been proven effective for modeling various physical phenomena including fluid dynamics [23, 24], solid mechanics [25, 26], and electromagnetic wave propagation [27]. Furthermore, PINNs can be seamlessly extended to solve real-world inverse problems. Notable applications include quantifying cardiovascular flow from visualization or sensor data [10, 28], spatial discovery of elastic modulus and Poisson's ratio [29, 30], and metamaterial design [31].

Although the wide range of applications in PINNs, it is still challenging to train an accurate PINN model. Some recent studies have focused on improving the performance of PINNs, such as using loss reweighting to promote the balancing of the training process and to improve the accuracy of tests [32-34]. Adaptive resampling methods like importance sampling [35], evolutionary sampling [36], and residual-based adaptive sampling [37], have also been utilized to achieve similar objectives. Moreover, new neural network architectures, including feature embedding [38] and adaptive activation functions [39], have been introduced to enhance the capacity of PINNs. However, the training of a PINN model requires a large number of collocation points to satisfy the differential equations and an extensive optimization iteration. Moreover, the number of collocation points and training iterations often increases with problem complexity, complicating the search for suitable hyperparameters.

In vanilla PINNs which are referred to as AD-PINNs in this paper, derivative terms are typically obtained through automatic differentiation (AD), However, it requires substantial GPU memory to store the computational graph of the network's derivatives during training. Moreover, AD derivatives cause the backpropagation chain to be longer, leading to inefficient training. Many researchers have proposed other effective approaches to overcome the above challenges. Sgouralis and Spiliopoulos [40] employed Monte Carlo methods to approximate second-order PDE derivatives. Chiu et al. [41] and Lim et al. [42] proposed finite difference schemes that use network output values to compute necessary differential operators for PINN training, offering robustness in sparse samples and efficiency compared to AD-PINN. Sharma et al. [43] and Xiao et al. [44] integrated meshless radial basis functions to compute spatial derivatives, theoretically handling arbitrary collocation point distributions. Navaneeth et al. [45] introduced a gradient-free physics-informed learning method based on random projections, successfully simulating fourth-order phase-field fracture problems. These methods alleviate the problems of computational inefficiency and low accuracy. However, derivative schemes may struggle with irregular geometries. Some methods are unsuitable for high-order PDE problems, and others necessitate additional collocation points beyond the boundary. Developing a robust and efficient differentiation method that resolves these challenges remains a critical pursuit.

The smoothing kernel is a crucial concept in smoothed particle hydrodynamics (SPH) [46] for characterizing the interaction relationship of different particles. This kernel function can be regarded as a substitute for the Dirac function, facilitating "smooth" spatial interactions. If regarding the collocation points of PINNs as the SPH particles, the derivative of PINNs can be replaced by an approximate derivative derived from the kernel differentiation formulas. These formulas are suitable for applications in domains with irregular geometries. Techniques derived by recovering particle consistency can approximate derivatives of any order without requiring virtual particles beyond the domain.

This paper introduces SK-PINN, a novel approach using a smoothing kernel function for derivative computation in place of AD in PINNs. SK-PINN utilizes a multi-layer perceptron (MLP) network layer with random Fourier feature embedding [47] for feature extraction from input data. It computes derivatives of collocation points through the smoothing kernel differentiation method. To address gradient imbalance among loss components during backpropagation, we employ an adaptive learning rate annealing algorithm proposed by Wang et al. [32] Additionally, we derive correction formulas for smoothing kernel derivatives of various orders and introduce an RKGM formula for simultaneous computation of derivatives of all orders. Compared to AD-PINN, SK-PINN demonstrates higher efficiency. We evaluate SK-PINN's performance by comparing it with AD-PINN and mAD-PINN (which only differs from SK-PINN in the derivatives) on inhomogeneous Poisson's equations. By using the neural tangent kernel theory [33, 48], we demonstrate that SK-PINN and AD-PINN have the same error convergence rate, but SK-PINN is more efficient. As the number of collocation points increases, the training cost ratio between AD-PINN and SK-PINN increases. Finally, we illustrate SK-PINN's superior performance across various forward and inverse problems.

This paper is structured as follows: Section 2 details the basic principles of physics-informed neural networks (PINNs) and discusses the enhancements applied in this study. Section 3 derives several smoothing kernel differentiation methods and compares their accuracy. Section 4 outlines the training process of SK-PINN and compares it with AD-PINN. Section 5 evaluates the performance of SK-PINN, AD-PINN, and mAD-PINN across various test cases. Final remarks are drawn in Section 6.

## 2. Physics-informed neural networks

Physics-informed neural networks (PINNs) combine the frameworks of deep learning with physical principles to improve accuracy. With this approach, we can use neural networks to approximate solutions to physical equations, while incorporating physical constraints during training to improve accuracy and generalization.

PINNs commonly employ deep neural network (DNN) architectures to represent the dynamical process [10, 28, 49, 50]. In the PINN model, $\hat{u}(x,t;\theta)$ is a prediction of the unknown variable $u$ within the spatiotemporal domain, where $x \in \Omega$ and $t \in (0, T]$. The dimension of the spatial domain $\Omega$ depends on the problem's dimensionality. The mapping relationship $(x, t) \to \hat{u}$ is determined by the network parameters $\theta$, which are optimized during training based on the PINN loss function.

Next, a general form of partial differential equations (PDEs) can be considered as

$$\begin{cases} \mathcal{N}_t[u(x,t)] + \mathcal{N}_x[u(x,t), \alpha] = 0, & x \in \Omega, t \in (0, T], \\ u(x, 0) = u_0(x), & x \in \Omega, \\ \mathcal{B}[u(x,t)] = 0, x \in \partial\Omega, & t \in (0, T]. \end{cases} \quad (1)$$

Here, $t$ represents time, $x \in \mathbb{R}^d$ represents the spatial coordinate. $\mathcal{N}_t[\cdot]$ and $\mathcal{N}_x[\cdot]$ represent differential operators in time and space, respectively. These differential operators can contain arbitrarily linear or nonlinear combinations of time-space derivatives. $\alpha$ represent the physical parameters of Eq. (1). $u_0(x)$ is the initial state. The boundary operator $\mathcal{B}[\cdot]$ can be used to represent different boundary conditions, such as Dirichlet, Neumann and Robin.

To obtain a solution $\hat{u}(x,t;\theta)$ satisfied all terms of Eq. (1), a comprehensive loss function can be defined as

$$\mathcal{L} = \lambda_{Data}\mathcal{L}_{Data} + \lambda_{DE}\mathcal{L}_{DE} + \lambda_{IC}\mathcal{L}_{IC} + \lambda_{BC}\mathcal{L}_{BC}. \tag{2}$$

Here,

- $\mathcal{L}_{Data}$ evaluates the fit to observable data, that is,

$$\mathcal{L}_{Data} = \frac{1}{n}\sum_{i=1}^{n}(u_i - \hat{u}_i)^2. \tag{3}$$

This feature of PINNs allows it to support solving inverse problems by incorporating experimental data.

- $\mathcal{L}_{DE}$ evaluates the residuals of the differential equations over the domain $\Omega \times (0,T]$.

$$\mathcal{L}_{DE} = \|\mathcal{N}_t[\hat{u}(\cdot;\theta)] + \mathcal{N}_x[\hat{u}(\cdot;\theta),\alpha]\|^2_{\Omega\times(0,T]}. \tag{4}$$

- $\mathcal{L}_{IC}$ evaluates the discrepancy at the initial states.

$$\mathcal{L}_{IC} = \|\hat{u}(\cdot,0;\theta) - u_0\|^2_{\Omega}. \tag{5}$$

- $\mathcal{L}_{BC}$ evaluates the boundary conditions.

$$\mathcal{L}_{BC} = \|\mathcal{B}[\hat{u}(\cdot;\theta)] - g(\cdot)\|^2_{\partial\Omega\times(0,T]}. \tag{6}$$

In Eq. (2), the loss weights $\lambda_*$ control the balance among these components and then affect the training outcomes of PINNs. $\mathcal{L}_{Data}$ (Eq. (3)) is defined as the mean square error between observed $u_i$ and predicted $\hat{u}_i$ values, where $n$ is the number of samples. $\mathcal{L}_{DE}$, $\mathcal{L}_{IC}$ and $\mathcal{L}_{BC}$ evaluate the residuals for differential equations, initial conditions and boundary conditions, respectively. It's noting that $\mathcal{L}_{Data}$ is omitted when solving forward differential equations.

The loss components associated with the equations are defined in a continuous spatiotemporal domain. During training, these are approximated using collocation points $D = \{(x_i, t_i)\}_{i=1}^{m}$, where $m$ denotes the number of collocation points. These points are typically sampled randomly from the problem domain using some methods such as equidistant grid spacing, randomized Latin hypercube sampling or importance sampling. Adaptive sampling [51-53] and causal sampling methods [54-56] have also been suggested to enhance the ability of PINNs.

We employ a strategy to optimize the PINN weights $\boldsymbol{\theta}$, where the Adam optimizer is used for pre-training and then $\boldsymbol{\theta}$ is further refined using the L-BFGS optimizer. Combining a first-order optimization method with a second-order optimization method (e.g., Adam + L-BFGS) can lead to higher convergences, as shown in [57]. In this paper, the learning rates used for Adam and L-BFGS are set to 0.001 and 1.0, respectively.

2.1. Neural tangent kernel

Generally, the neural tangent kernel (NTK) theory provides a theoretical perspective to analyze the training of DNN, which was first proposed by Jacot et al. [48]. In the NTK theory, the functional regression is considered by using a deep fully-connected network $f(\boldsymbol{x}, \boldsymbol{\theta})$, where the weights $\boldsymbol{\theta}$ are initialized from a Gaussian distribution $\mathcal{N}(0,1)$. Given dataset $\{\boldsymbol{X}_{train}, \boldsymbol{Y}_{train}\}$, where $\boldsymbol{X}_{train} = \{\boldsymbol{x}_i\}_{i=1}^N$ and $\boldsymbol{Y}_{train} = \{y_i\}_{i=1}^N$. If the network parameters $\boldsymbol{\theta}$ are trained with a very small learning rate $\eta$ to minimize the loss $\mathcal{L}(\boldsymbol{\theta}) = \frac{1}{2}\sum_{i=1}^N (f(\boldsymbol{x}_i, \boldsymbol{\theta}) - y_i)^2$, this training yields

$$f(\boldsymbol{X}_{train}, \boldsymbol{\theta}) = (\boldsymbol{I} - e^{-\boldsymbol{K}t'}) \cdot \boldsymbol{Y}_{train}, \tag{7}$$

where $t'$ represents the trained time, specifically referring to the number of training iterations or steps. According to [48], the neural tangent kernel $\boldsymbol{K}$ is defined as

$$\boldsymbol{K}_{ij} = \boldsymbol{K}(\boldsymbol{x}_i, \boldsymbol{x}_j) = \langle \frac{\partial f(\boldsymbol{x}_i, \boldsymbol{\theta})}{\partial \boldsymbol{\theta}}, \frac{\partial f(\boldsymbol{x}_j, \boldsymbol{\theta})}{\partial \boldsymbol{\theta}} \rangle. \tag{8}$$

The NTK theory asserts that the kernel $\boldsymbol{K}$ converges to a deterministic kernel under gradient flow with an infinitesimally small learning rate. This property remains unchanged during training even as the network width grows to infinity.

However, Eq (7) can be derived

$$\boldsymbol{Q}^{\mathrm{T}}[f(\boldsymbol{X}_{train}, \boldsymbol{\theta}) - \boldsymbol{Y}_{train}] = -e^{-\boldsymbol{\Lambda}t'}\boldsymbol{Q}^{\mathrm{T}}\boldsymbol{Y}_{train}. \tag{9}$$

Here, $\boldsymbol{\Lambda}$ is a diagonal matrix of eigenvalues $\lambda_i$ and $\boldsymbol{\Lambda} = \boldsymbol{Q}^{\mathrm{T}}\boldsymbol{K}\boldsymbol{Q}$. $\boldsymbol{Q}$ is an orthogonal matrix with eigenvectors $\boldsymbol{q}_i$. From Eq. (9), the convergence rate of $\boldsymbol{q}_i^{\mathrm{T}}[f(\boldsymbol{X}_{train}, \boldsymbol{\theta}) - \boldsymbol{Y}_{train}]$ is influenced by the eigenvalue $\lambda_i$, highlighting the network's bias towards learning along eigendirections with larger eigenvalues. Detailed analysis by Cao et al. [58] reveals that for conventional fully-connected neural networks, NTK eigenvalues decrease monotonically with increasing eigenfunction frequency, leading to slower convergence of high-frequency components in the objective function [59, 60], known as "spectral bias."

Moreover, Wang et al. [33] found that spectral bias also occurs in PINN. When a DNN $u(x; \boldsymbol{\theta})$ is used approximate the solution of the following PDE $\begin{cases} \mathcal{N}_x[u(x)] = f(x), & x \in \Omega, \\ \mathcal{B}[u(x)] = g(x), & x \in \partial\Omega. \end{cases}$ the loss function is defined as $\mathcal{L}(\boldsymbol{\theta}) = \frac{1}{2}\sum_{i=1}^{N_b}|\mathcal{B}[u(x_b^i; \boldsymbol{\theta})] - g(x_b^i)|^2 + \frac{1}{2}\sum_{i=1}^{N_r}|\mathcal{N}_x[u(x_r^i; \boldsymbol{\theta})] - f(x_r^i)|^2$. Here, $N_r$ and $N_b$ represent the in-domain coordinate dataset $\{x_r^i\}_{i=1}^{N_r}$ and the Dirichlet boundary dataset $\{x_b^i, g(x_b^i)\}_{i=1}^{N_b}$, respectively. A series of derivations leads to

$$Q^{\mathrm{T}} \begin{pmatrix} \mathcal{B}[u(x_b^i; \boldsymbol{\theta})] - g(x_b^i) \\ \mathcal{N}_x[u(x_r^i; \boldsymbol{\theta})] - f(x_r^i) \end{pmatrix} = -e^{-\Lambda t'} Q^{\mathrm{T}} \begin{pmatrix} g(x_b^i) \\ f(x_r^i) \end{pmatrix}. \tag{10}$$

Similarly, the neural tangent kernel of PINNs converges to a deterministic kernel under appropriate conditions and remains unchanged during training via gradient descent with an infinitesimally small learning rate. The fully-connected PINN not only suffers from spectral bias but also exhibits significant differences in the convergence rates of various components within its loss function.

2.2. Random Fourier feature embedding

To handle the spatial bias, Tancik et al. [47] proposed a random Fourier feature embedding by mapping input coordinates to high-frequency signals before passing through the DNN. This encoding is defined as

$$\gamma(x) = \begin{bmatrix} \cos(\boldsymbol{B}x) \\ \sin(\boldsymbol{B}x) \end{bmatrix}. \tag{11}$$

Here $\boldsymbol{B} \in \mathbb{R}^{l \times d}$ and each entry is sampled from the Gaussian distribution $\mathcal{N}(0, \sigma^2)$, where $\sigma$ is a user-specified hyperparameter. This simple approach has been shown to significantly improve the performance of PINNs in approximating sharp gradients and complex solutions [38]. The scaling factor $\sigma$ has a significant effect on the performance of the neural network. This hyperparameter directly controls the frequency of $\boldsymbol{\gamma}$, which in turn controls the feature space of the NTK, thus biasing the network towards learning certain band-limited signals. Specifically, lower encoding frequencies may lead to blurred predictions, while higher encoding frequencies may introduce salt-and-pepper artifacts. Ideally, $\sigma$ should be chosen such that the bandwidth of the NTK matches the bandwidth of the target signal. This approach not only speeds up the training convergence, but also improves the prediction accuracy. In this paper, we follow the setting $\sigma \in [1,10]$ of Wang et al. [61].

2.3. Adaptive loss balancing

It is crucial to select the weighting coefficients for different loss components in the PINNs. Manual weight selection is impractical because optimal weights can vary significantly across different problems. Furthermore, there is no universally applicable empirical formulation that can be transferred to various PDEs, especially when the solution is unknown and no validation dataset is available for hyperparameter tuning.

In this study, we adopt the adaptive learning rate annealing algorithm proposed by Wang et al. [61]), which automatically balances losses during training. The formula ensures equal gradient norms across the weighted loss terms. Specifically, the weighting coefficients are obtained through the following

$$\left\| \lambda_{BC} \frac{\partial \mathcal{L}_{BC}}{\partial \boldsymbol{\theta}} \right\| = \left\| \lambda_{DE} \frac{\partial \mathcal{L}_{DE}}{\partial \boldsymbol{\theta}} \right\| = \left\| \frac{\partial \mathcal{L}_{BC}}{\partial \boldsymbol{\theta}} \right\| + \left\| \frac{\partial \mathcal{L}_{DE}}{\partial \boldsymbol{\theta}} \right\|. \tag{12}$$

This approach effectively ensures that the gradients of each weighted loss are balanced, preventing the model from biasing towards minimizing specific terms during training. These updates can be scheduled every one hundred or one thousand iterations of a gradient descent loop or at a frequency specified by the user. Consequently, any additional computational overhead associated with these updates is minimal, particularly when updates are infrequent.

## 3. Smoothed kernel theory

Smoothed particle hydrodynamics (SPH) is a well-established mesh-free method widely utilized in various fields of mechanics [5, 46]. The core of the SPH method is kernel approximation, which uses smoothing kernel functions to represent a function and its derivatives. This section introduces the properties of the kernel function and explains how to approximate the derivatives of field functions using smoothing kernel functions. Subsequently, several commonly used correction formulas are provided to enhance the accuracy of derivative approximations. Finally, RKGM is proposed to calculate all spatial derivatives simultaneously. The accuracy and characteristics of these formulas will be presented in Section 3.2.

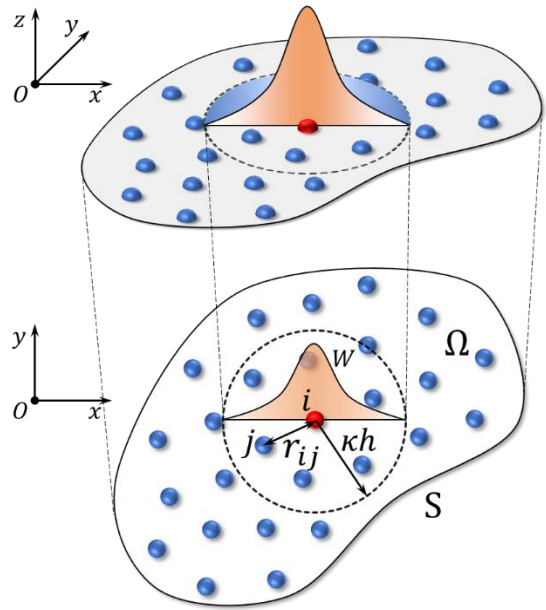

**Fig. 1.** SPH particle approximations in a two-dimensional problem domain $\Omega$ with a surface S. $W$ is the smoothing kernel function used to approximate the field variables at particle $i$ by averaging summations over particles $j$ within the support domain, defined by a cut-off distance of $\kappa h$.

First, an arbitrary function $f(x)$ satisfies the following

$$f(x) = \int_\Omega f(x')\delta(x-x')\mathrm{d}x', \tag{13}$$

where $f$ represents a function of the position vector $x$, and the Dirac function $\delta(x-x')$ is defined as

$$\delta(x-x') = \begin{cases} \infty, x = x' \\ 0, x \neq x' \end{cases} \text{ and } \int_\Omega \delta(x-x')\mathrm{d}x' = 1. \tag{14}$$

In Eq. (13), $\Omega$ is the volume of the integral that contains $x$. Eq. (13) implies that a function can be represented in an integral form. Since the Dirac function is used, the integral representation in Eq. (13) is exact and rigorous, as long as $f(x)$ is defined and continuous in $\Omega$.

The Dirac function $\delta(x-x')$ has only "point" support, which means Eq. (13) cannot be used for establishing discrete numerical models. If replacing the Dirac function $\delta(x-x')$ by a smoothing kernel function $W(x-x',h)$ with a finite spatial length $h$, the kernel approximation of $f(x)$ denoted as $\langle f(x) \rangle$ becomes

$$\langle f(x) \rangle = \int_\Omega f(x')W(x-x',h)\mathrm{d}x', \tag{15}$$

where $h$ is the smooth length defining the influence or support area of the smoothing kernel function $W$. It is important to note that the integral representation shown in Eq. (15) can only approximate $f(x)$, as long as $W$ is not the Dirac function.

A smoothing kernel function $W$ is typically chosen to be an even function and possesses the following properties

$$\int_\Omega W(x-x',h)\mathrm{d}x' = 1, \tag{16}$$

$$\lim_{h \to 0} W(x-x',h) = \delta(x-x'), \tag{17}$$

$$W(x-x',h) = 0, |x-x'| > \kappa h, \tag{18}$$

where $\kappa$ is a scaling factor, and $\kappa h$ defines the effective (non-zero) area of $W$. This effective area is often referred to as the support domain of the smoothing function for a point at $x$. By using this compact condition, integration over the entire problem domain is localized to the support domain of the smoothing function. Therefore, the integration domain $\Omega$ can be considered the same as the support domain, as illustrated in Fig. 1. Consequently, the kernel approximation of the derivatives can be derived (for a detailed derivation see [46])

$$\langle \nabla f(\boldsymbol{x}) \rangle = - \int_\Omega f(\boldsymbol{x}') \nabla W(\boldsymbol{x} - \boldsymbol{x}', h) \mathrm{d}\boldsymbol{x}'. \tag{19}$$

The differential operation on a function is transformed into a differential operation on the kernel function. Essentially, the spatial gradient of a function can be computed using the values of the function and the derivatives of the smoothing kernel function, rather than relying on the derivatives of the function itself.

In the SPH, a set of particles (typically arbitrarily distributed, as illustrated in Fig. 1 in a two-dimensional domain) is employed to represent the problem domain and approximate the field variables. These particles can initially be generated as centered particles using existing mesh generation tools. The state of the system is characterized by these particles, each associated with specific field properties. They serve not only for integration, interpolation, or differencing but also for material representation. The volume of a subsection is consolidated into the corresponding particle, meaning each particle $i$ is assigned a fixed lumped volume $\Delta V_i$ without a predefined shape. These particles can either remain fixed in the Eulerian frame or move within the Lagrangian frame.

Once the computational domain is represented by a finite number of particles, the continuous form of kernel approximation is expressed in Eqs. (15) and (19) can be discretized into a summation over neighboring particles, as follows

$$\langle f(\boldsymbol{x}_i) \rangle = \sum_{j=1}^{N} f_j W_{ij} \Delta V_j, \tag{20}$$

$$\langle \nabla f(\boldsymbol{x}_i) \rangle = - \sum_{j=1}^{N} f_j \boldsymbol{\nabla}_j W_{ij} \Delta V_j = \sum_{j=1}^{N} f_j \boldsymbol{\nabla}_i W_{ij} \Delta V_j, \tag{21}$$

where $N$ is the total number of particles in the support domain of particle $i$, and the radius of the support domain is $\kappa h$. To simplify the formula, we replace $f_j$ with $f(x_j)$, $x_{ij}$ with $x_i - x_j$ and $W_{ij}$ with $W(x_i - x_j, h)$. In this paper we refer to Eq. (21) as the SPH formula.

Regarding the selection of the smoothing kernel function, a commonly used choice is the cubic B-spline function, originally proposed by Monaghan and Lattanzio [62]. The cubic B-spline kernel function $W(R, h)$ is defined as

$$W(R, h) = \alpha_d \begin{cases} \frac{2}{3} - R^2 + \frac{1}{2}R^3, & 0 \leq R < 1, \\ \frac{1}{6}(2 - R)^3, & 1 \leq R < 2, \\ 0, & R \geq 2. \end{cases} \quad (22)$$

In one-, two- and three-dimensional space, $\alpha_d$ is $1/h$, $15/7\pi h^2$ and $3/2\pi h^3$, respectively. The cubic spline function is widely regarded as the most commonly used kernel function in SPH. Because it closely resembles the Gaussian function but with a narrower compact support.

3.1. Improved formula based on Taylor series expansion

Applying Eq. (21) presents several challenges. Firstly, points near boundaries may encounter a lack of neighboring points (as depicted in Fig. 2), leading to inaccuracies in derivative calculations. Additionally, the non-uniform distribution of points within the support domain can introduce errors from the integral-to-discrete conversion. To address these issues, several correction formulas are derived based on Taylor series expansions, and a regenerative kernel gradient correction technique is proposed.

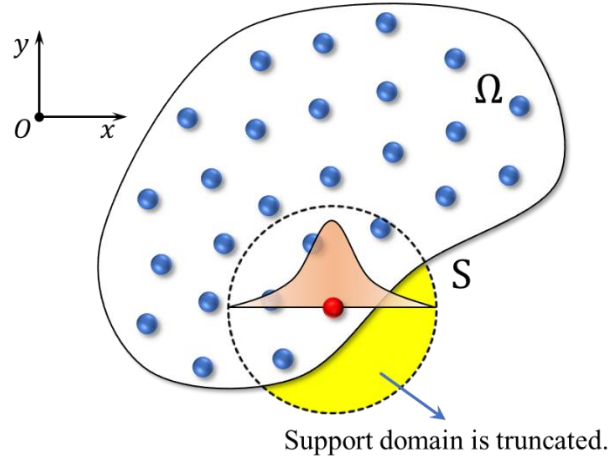

Support domain is truncated.

**Fig. 2.** Particle approximation for a particle whose support domain is truncated by the boundary.

3.1.1. Corrective smoothed particle method

By implementing a Taylor series expansion around a nearby point $x_i$, a sufficiently smooth function $f(x)$ can be expressed as

$$f(x) = f_i + \sum_{\alpha=1}^{d}(x^\alpha - x_i^\alpha)f_{i,\alpha} + \sum_{\alpha=1}^{d}\sum_{\beta=1}^{d}\frac{(x^\alpha - x_i^\alpha)(x^\beta - x_i^\beta)}{2!}f_{i,\alpha\beta} + \cdots, \tag{23}$$

where $d$ represents the dimension of $x$. $\alpha$ and $\beta$ represent the dimension indices. $f_{i,\alpha}$ and $f_{i,\alpha\beta}$ denote the first-order and second-order derivatives of $f(x_i)$ in different dimensions, respectively.

To simplify, we remove terms in Eq. (23) involving second-order derivatives and higher, retaining the term related to the dimension index $\gamma$ for the first-order terms. Then, both sides of (23) are multiplied by the kernel gradient $W_{i,\gamma}$ and integrated over the entire domain to obtain

$$\int_\Omega f(x)W_{i,\gamma}\mathrm{d}x = f_i\int_\Omega W_{i,\gamma}\mathrm{d}x + f_{i,\gamma}\int_\Omega (x^\gamma - x_i^\gamma)W_{i,\gamma}\mathrm{d}x, \tag{24}$$

where $W_{i,\gamma}$ represents $\partial W(x_i - x, h)/\partial x_i^\gamma$. After discretizing (24), the following formula can be derived

$$f_{i,\gamma} = \frac{\sum_j[f_j - f_i]W_{ij,\gamma}\Delta V_j}{\sum_j\left(x_j^\gamma - x_i^\gamma\right)W_{ij,\gamma}\Delta V_j}. \tag{25}$$

Here, $W_{ij,\gamma}$ represents $\partial W(\boldsymbol{x}_i - \boldsymbol{x}_j, h)/\partial x_i^{\gamma}$. This is known as the corrective smoothed particle method [63], and we refer to (25) as the CSPH formula in this paper.

3.1.2. Kernel gradient correction

The second-order derivatives and higher terms in Eq. (23) are removed. Then, both sides of Eq. (23) are multiplied by the kernel gradient $W_{i,\gamma}$ and integrated over the entire domain to obtain

$$\int_{\Omega} f(\boldsymbol{x}) W_{i,\gamma} \mathrm{d}\boldsymbol{x} = f_i \int_{\Omega} W_{i,\gamma} \mathrm{d}\boldsymbol{x} + \sum_{\alpha=1}^{d} f_{i,\gamma} \int_{\Omega} (x^{\alpha} - x_i^{\alpha}) W_{i,\gamma} \mathrm{d}\boldsymbol{x}. \tag{26}$$

Eq. (26) is replaced by the discrete equation and then derived to obtain [64]

$$f_{i,\gamma} = \left[\sum_j (\boldsymbol{x}_j - \boldsymbol{x}_i) \nabla_i^{\mathrm{T}} W_{ij} \Delta V_j\right]^{-1} \sum_j (f_j - f_i) W_{ij,\gamma} \Delta V_j. \tag{27}$$

As an example of a two-dimensional problem, we use $\boldsymbol{x} = [x, y]^{\mathrm{T}}$ to represent different dimensional coordinates. Eq. (27) can be written as

$$\begin{bmatrix} f_{i,x} \\ f_{i,y} \end{bmatrix} = \begin{bmatrix} \sum_j x_{ji} W_{ij,x} \Delta V_j & \sum_j y_{ji} W_{ij,x} \Delta V_j \\ \sum_j x_{ji} W_{ij,y} \Delta V_j & \sum_j y_{ji} W_{ij,y} \Delta V_j \end{bmatrix}^{-1} \begin{bmatrix} \sum_j f_{ji} W_{ij,x} \Delta V_j \\ \sum_j f_{ji} W_{ij,y} \Delta V_j \end{bmatrix}, \tag{28}$$

where $x_{ji}$ and $y_{ji}$ denotes $(x_j - x_i)$ and $(y_j - y_i)$, respectively. $f_{ji}$ denotes $f(\boldsymbol{x}_j) - f(\boldsymbol{x}_i)$. In this paper we refer to Eq. (27) as the KGC formula.

3.1.3. Finite particle method

The finite particle method [65, 66] selects a set of basis functions $\boldsymbol{\varphi}_M$. Then, both sides of (23) are multiplied by each function $\boldsymbol{\varphi}_M$ and integrated over the entire domain to obtain

$$\int_\Omega f(\boldsymbol{x})\boldsymbol{\varphi}_M \mathrm{d}\boldsymbol{x} = f_i \int_\Omega \boldsymbol{\varphi}_M \mathrm{d}\boldsymbol{x} + \sum_{\alpha=1}^{d} f_{i,\alpha} \int_\Omega (x^\alpha - x_i^\alpha)\boldsymbol{\varphi}_M \mathrm{d}\boldsymbol{x}$$
$$+ \sum_{\alpha=1}^{d}\sum_{\beta=1}^{d} f_{i,\alpha\beta} \int_\Omega \frac{(x^\alpha - x_i^\alpha)(x^\beta - x_i^\beta)}{2!}\boldsymbol{\varphi}_M \mathrm{d}\boldsymbol{x} + \cdots. \tag{29}$$

The more basis functions $\boldsymbol{\varphi}_M$ takes, the higher order terms Eq. (29) can retain, which means the higher the accuracy of the derivatives. We take $\boldsymbol{\varphi}_M = [W_i, W_{i,x}, W_{i,y}]^\mathrm{T}$ to obtain

$$\begin{bmatrix} f_i \\ f_{i,x} \\ f_{i,y} \end{bmatrix} = \begin{bmatrix} \sum_j W_{ij}\Delta V_j & \sum_j x_{ji} W_{ij}\Delta V_j & \sum_j y_{ji} W_{ij}\Delta V_j \\ \sum_j W_{ij,x}\Delta V_j & \sum_j x_{ji} W_{ij,x}\Delta V_j & \sum_j y_{ji} W_{ij,x}\Delta V_j \\ \sum_j W_{ij,y}\Delta V_j & \sum_j x_{ji} W_{ij,y}\Delta V_j & \sum_j y_{ji} W_{ij,y}\Delta V_j \end{bmatrix}^{-1} \begin{bmatrix} \sum_j f_{ji} W_{ij}\Delta V_j \\ \sum_j f_{ji} W_{ij,x}\Delta V_j \\ \sum_j f_{ji} W_{ij,y}\Delta V_j \end{bmatrix}. \tag{30}$$

In this paper, we refer to Eq. (30) as the FPM formula.

3.1.4. Reproducing kernel gradient method

The reproducing kernel gradient method is based on the idea of the reproducing kernel particle method (RKPM) [67], which introduces a set of reproducing kernel functions $\widetilde{W}_i$. To illustrate the construction of $\widetilde{W}_i$, we multiply both sides of Eq. (23) by $\widetilde{W}_i$ and integrate over the entire computational domain to obtain

$$\int_\Omega f(\boldsymbol{x})\widetilde{W}_i \mathrm{d}\boldsymbol{x} = f_i \int_\Omega \widetilde{W}_i \mathrm{d}\boldsymbol{x} + \sum_{\alpha=1}^{d} f_{i,\alpha} \int_\Omega (x^\alpha - x_i^\alpha)\widetilde{W}_i \mathrm{d}\boldsymbol{x}$$
$$+ \sum_{\alpha=1}^{d}\sum_{\beta=1}^{d} f_{i,\alpha\beta} \int_\Omega \frac{(x^\alpha - x_i^\alpha)(x^\beta - x_i^\beta)}{2!}\widetilde{W}_i \mathrm{d}\boldsymbol{x} + \cdots, \tag{31}$$

where the reproducing kernel $\widetilde{W}_i$ is

$$\widetilde{W}_i = C_i W_i. \tag{32}$$

Here, $W_i$ represents $W(\boldsymbol{x}_i - \boldsymbol{x}, h)$ and $C_i$ is a linear combination of a set of polynomial basis functions

$$\begin{cases} C_i = \boldsymbol{H}_i^{\mathrm{T}} \boldsymbol{b}_i, \\ \boldsymbol{H}_i^{\mathrm{T}} = \left[1, x^\alpha - x_i^\alpha, (x^\alpha - x_i^\alpha)(x^\beta - x_i^\beta), \ldots \right], \\ \boldsymbol{b}_i^{\mathrm{T}} = \left[b_i^0, b_i^\alpha, b_i^{\alpha\beta}, \ldots \right], \end{cases} \quad (33)$$

where $\boldsymbol{H}_i^{\mathrm{T}}$ represents a set of complete polynomial basis functions and $\boldsymbol{b}_i^{\mathrm{T}}$ is the unknown coefficient determined by the distribution position of neighboring points. The number of terms included in $\boldsymbol{H}_i^{\mathrm{T}}$ corresponds to those retained by Eq. (31).

Clearly, increasing the number of terms in $\boldsymbol{H}_i^{\mathrm{T}}$ theoretically improves the accuracy of $\int_\Omega f(\boldsymbol{x})\widetilde{W}_i \mathrm{d}\boldsymbol{x}$. If we aim to approximate $f_i$ using $\int_\Omega f(\boldsymbol{x})\widetilde{W}_i \mathrm{d}\boldsymbol{x}$, then we have

$$\int_\Omega \boldsymbol{H}_i \boldsymbol{H}_i^{\mathrm{T}} \boldsymbol{b}_i W_i \mathrm{d}\boldsymbol{x} = \boldsymbol{H}_0^{\mathrm{T}}. \quad (34)$$

Here, $\boldsymbol{H}_0^{\mathrm{T}} = \left[1, 0^\alpha, 0^{\alpha\beta}, \ldots \right]$, $0^\alpha$ and $0^{\alpha\beta}$ indicate that the number of zeros taken corresponds to $x^\alpha - x_i^\alpha$ and $(x^\alpha - x_i^\alpha)(x^\beta - x_i^\beta)$, respectively. After discretizing Eq. (34), the following formula can be derived

$$\boldsymbol{b}_i = \left[\sum_j \boldsymbol{H}_i \boldsymbol{H}_i^{\mathrm{T}} W_{ji} \Delta V_j \right]^{-1} \boldsymbol{H}_0^{\mathrm{T}}. \quad (35)$$

Therefore, $f_i$ can be expressed as

$$f_i = \sum_j f(\boldsymbol{x}_j) \boldsymbol{H}_i \boldsymbol{b}_i^T W_{ji} \Delta V_j. \quad (36)$$

In the RKPM, $f_{i,\alpha}$ is obtained by the derivation of the chain rule of Eq. (36). In this work, $\boldsymbol{H}_0^{\mathrm{T}}$ is changed so that $\int_\Omega f(\boldsymbol{x})\widetilde{W}_i \mathrm{d}\boldsymbol{x}$ approximates $f_{i,\alpha}$ and $f_{i,\alpha\beta}$. As an example of a two-dimensional problem, the terms higher the second-order in Eq. (31) are removed and taking the

$$\begin{cases} \boldsymbol{H}_i^{\mathrm{T}} = [1, x - x_i, y - y_i, (x - x_i)^2, (x - x_i)(y - y_i), (y - y_i)^2], \\ \boldsymbol{b}_i^{\mathrm{T}} = [b_i^0, b_i^1, b_i^2, b_i^{11}, b_i^{12}, b_i^{22}]. \end{cases} \quad (37)$$

Then, the following formula is obtained

$$[f_i, f_{i,x}, f_{i,y}, f_{i,xx}, f_{i,xy}, f_{i,yy}]^{\mathrm{T}} = \sum_j f(\boldsymbol{x}_j) \boldsymbol{H}_i \left[\sum_j \boldsymbol{H}_i \boldsymbol{H}_i^{\mathrm{T}} W_{ji} \Delta V_j \right]^{-1} \boldsymbol{\Lambda}_0^{\mathrm{T}} W_{ji} \Delta V_j, \quad (38)$$

where $\mathbf{\Lambda}_0^T = diag(1,1,1,2,1,2)$. Eq. (38) implies that the different derivatives can be computed separately and do not significantly increase the computing time. In this paper, we refer to Eq. (38) as the RKGM formulation with second-order accuracy.

## 3.2. Errors of different derivative formulas

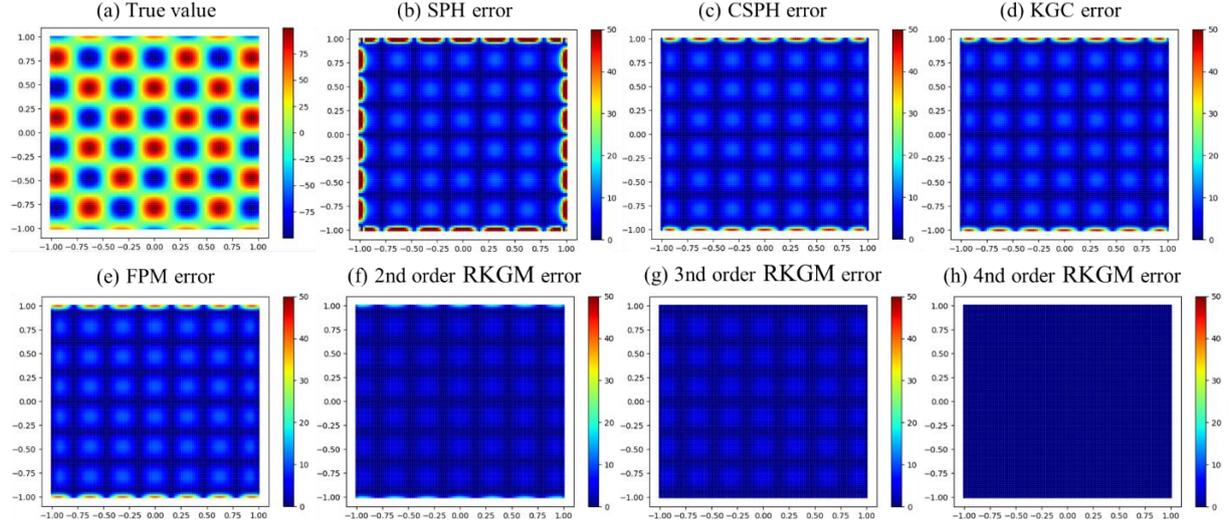

**Fig. 3.** Color map of calculation errors for different derivative formulas.

**Table 1.** Maximum absolute errors for different derivative formulas.

| Derivative formulas | Maximum absolute error |
| --- | --- |
| SPH | 79.951 |
| CSPH | 47.021 |
| KGC | 47.021 |
| FPM | 40.155 |
| 2nd order RKGM | 17.298 |
| 3nd order RKGM | 4.4443 |
| 4nd order RKGM | 0.3536 |

This section demonstrates the differences among derivative formulas in Section 3.1 by evaluating the second-order partial derivatives $\frac{\partial^2 u}{\partial x \partial y}$ for $u = \sin(10x) + \cos(10y)$ over the interval $x \in [-1,1], y \in [-1,1]$. Fig. 3a shows the true values of $\frac{\partial^2 u}{\partial x \partial y}$ and Figs. 3b-h illustrate the derivative errors using the SPH, CSPH, KGC, FPM, and various orders of the RKGM formulas, respectively. The maximum absolute errors are summarized in Table 1.

Notably, the SPH, CSPH, KGC, and FPM formulas require multiple steps to obtain second-order derivatives, whereas the RKGM formula can compute $\frac{\partial^2 u}{\partial x \partial y}$ directly in one step. Fig. 3b shows that the boundary derivatives obtained by the SPH formulation are significantly inaccurate. the CSPH, KGC and FPM formulations perform slightly better but still have errors at the boundary.

In real-time particle position update scenarios, the CSPH, KGC, and FPM formulas have an advantage due to their non-modification of the smoothing kernel function. However, second-order or higher-order RKGM formulas can lead to more accurate results in the SK-PINN.

The RKGM formula was chosen as the method for calculating the derivatives of SK-PINN, given its ability to achieve arbitrary order accuracy and to compute different order derivatives in a single step. In general, it is used the second-order accuracy of the RKGM formula.

## 4. SK-PINN

This section outlines the solution procedure of SK-PINN for solving a general steady-state PDE, as illustrated in Algorithm 1. Time-dependent PDEs require careful consideration of causality [56, 61, 68]. The derivative formulas derived in Section 3 focus on spatial derivatives, so time-dependent terms are excluded in this paper.

**Algorithm 1** SK-PINN Computational procedure

1. Non-dimensionalize the PDE system $\begin{cases} \mathcal{N}_x[u(x)] = f(x), & x \in \Omega, \\ \mathcal{B}[u(x)] = g(x), & x \in \partial\Omega. \end{cases}$

2. In the computational domain, the coordinate dataset $\{x_r^i\}_{i=1}^{N_r}$ and the boundary dataset $\{x_b^i, g(x_b^i)\}_{i=1}^{N_b}$ are obtained. The solution $u(x; \boldsymbol{\theta})$ is approximated using a DNN with random Fourier feature embeddings, utilizing the tanh activation function and Glorot [69] initialization.

3. The weighted loss function is defined as follows
$$\mathcal{L} = \lambda_{DE}\mathcal{L}_{DE} + \lambda_{BC}\mathcal{L}_{BC}.$$

4. $\mathcal{L}_{DE}$ and $\mathcal{L}_{BC}$ follow the (4) and (6) described in Section 2, where the $\mathcal{N}_x[\cdot]$ operator is computed using (38). The initial values for $\lambda_{DE}$ and $\lambda_{BC}$ are set to 1.0.

5. Update the parameters $\boldsymbol{\theta}$ using the Adam optimizer for $N_1$ iterations.

6. **For** $n = 1, \ldots, N_1$ **do**
7.     **If** $n \bmod f = 0$ **then**
8.         (a) Calculate the new weight coefficients $\hat{\boldsymbol{\lambda}} = (\hat{\lambda}_{DE}, \hat{\lambda}_{BC})$ according to (12).
9.         (b) Update the global weights $\boldsymbol{\lambda}_n = (\lambda_{DE}, \lambda_{BC})$ using a moving average of the form
$$\boldsymbol{\lambda}_{n+1} = \alpha \boldsymbol{\lambda}_n + (1-\alpha)\hat{\boldsymbol{\lambda}},$$
        where the parameter $\alpha$ determines the balance between $\boldsymbol{\lambda}_n$ and $\boldsymbol{\lambda}_{n-1}$.
10.     **End if**
11.     (c) Update the parameters $\boldsymbol{\theta}$
12. **End for**
13. Finally, update the parameters $\boldsymbol{\theta}$ using L-BFGS optimizer for $N_2$ iterations.
14. **For** $n = N_1 + 1, \ldots, N_1 + N_2$ **do**
15.     (a) Update the parameters $\boldsymbol{\theta}$
16. **End for**

The recommended default values for hyper-parameters are as follows: $f = 1000$, $\alpha = 0.9$. The learning rates for Adam and L-BFGS are 1e-3 and 1.0, respectively.

The SK-PINN program is implemented using Python and the open-source deep learning framework PyTorch 1.12.0. The complete source code of the proposed SK-PINN method in this work is available on GitHub at https://github.com/pcl-china/SK-PINN. Unless otherwise specified, training and testing of the network were conducted on a microcomputer equipped with a 12th Gen Intel Core i9-12900K CPU and an 8 GB NVIDIA GeForce RTX 3070 Ti GPU.

## 4.1. Comparison of training efficiency of SK-PINNs vs AD-PINNs

In this section, the characteristics of SK-PINN are demonstrated from several perspectives. First, from the neural tangent kernel (NTK) perspective, AD-PINN and SK-PINN total training error convergence rates are in the same order of magnitude. Secondly, considering a denser distribution of collocation points, we observe that SK-PINN achieves higher computational efficiency compared to AD-PINN as the number of collocation points increases. Under identical training configurations, SK-PINN can reduce training time significantly.

Next, the relationship between the average relative errors of AD-PINN and SK-PINN over training time is compared when solving the inhomogeneous Poisson's equation by using the correction methods described in Section 2. Finally, we propose an SK-AD training strategy aimed at rapidly reducing errors.

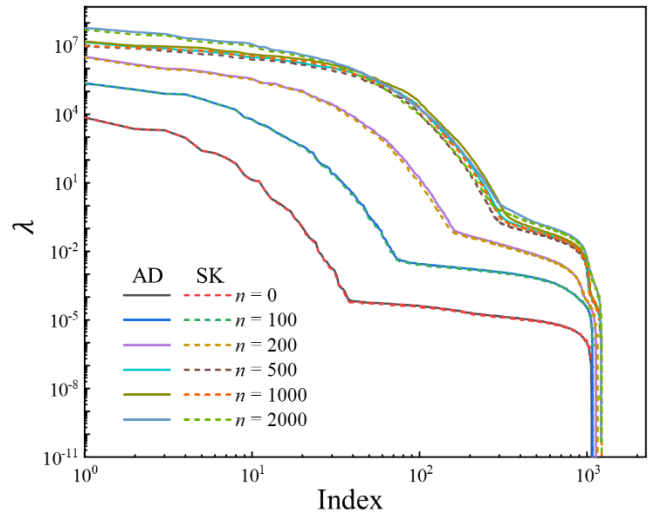

**Fig. 4.** Distribution of NTK eigenvalues of SK-PINN and AD-PINN for the same training.

We compare the RKGM-based differentiation technique (called SK technique) with AD-PINN's AD technique in the inhomogeneous Poisson's equation. AD-PINN employs automatic differentiation to compute derivative terms within differential equations, while SK-PINN utilizes the RKGM formula for this purpose. Both techniques allow us to analyze the neural tangent kernel (NTK) eigenvalue distribution during PINN training.

Figure 4 illustrates the comparison. The $x$-axis represents the NTK eigenvalue index, and the $y$-axis represents the eigenvalue magnitude. The solid line represents the NTK eigenvalue distribution of the AD, while the dashed line represents the distribution of the SK. $n$ denotes the current training iteration. NTK eigenvalues are sorted in descending order. At the same training iterations, the NTK eigenvalue distributions of both AD and SK exhibit comparable magnitudes. This suggests that AD-PINN and SK-PINN achieve a similar rate of convergence in total training error. Furthermore, the NTK converges rapidly with increasing training iterations, aligning with the findings of Wang et al. [33] that the NTK of PINN converges towards a deterministic kernel.

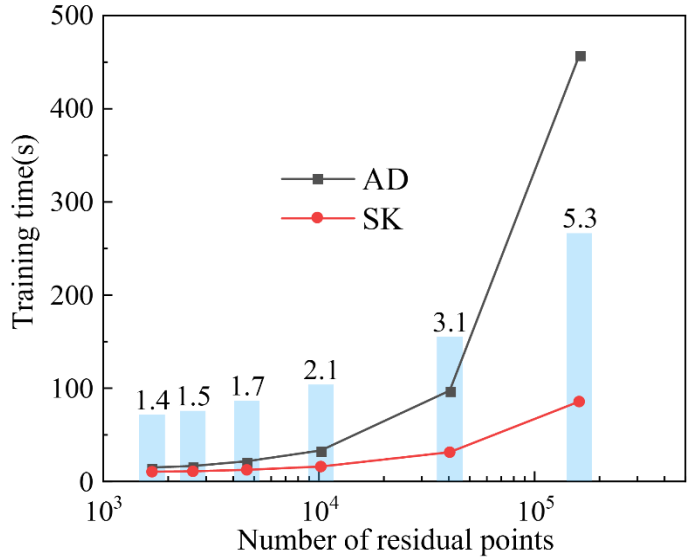

**Fig. 5.** Comparison of training time between SK-PINN and AD-PINN using varying numbers of collocation points.

Figure 5 illustrates the comparison of training time between SK-PINN and AD-PINN when solving the inhomogeneous Poisson's problem with varying numbers of collocation points. Here, we initially pre-train using Adam for 1000 iterations, followed by an additional 1000 iterations using L-BFGS. The black dotted line represents the training time for AD-PINN, the red dotted line represents the training time for SK-PINN, and the light blue bars indicate the ratio of training time between AD-PINN and SK-PINN. SK-PINN consistently exhibits shorter training time compared to AD-PINN. Moreover, SK-PINN demonstrates significantly higher training efficiency than AD-PINN as the number of collocation points increases. This underscores SK-PINN's distinct advantage in tackling problems requiring numerous collocation points. For instance, SK-PINN notably reduces training time when computing three-dimensional problems that require a large number of spatial points.

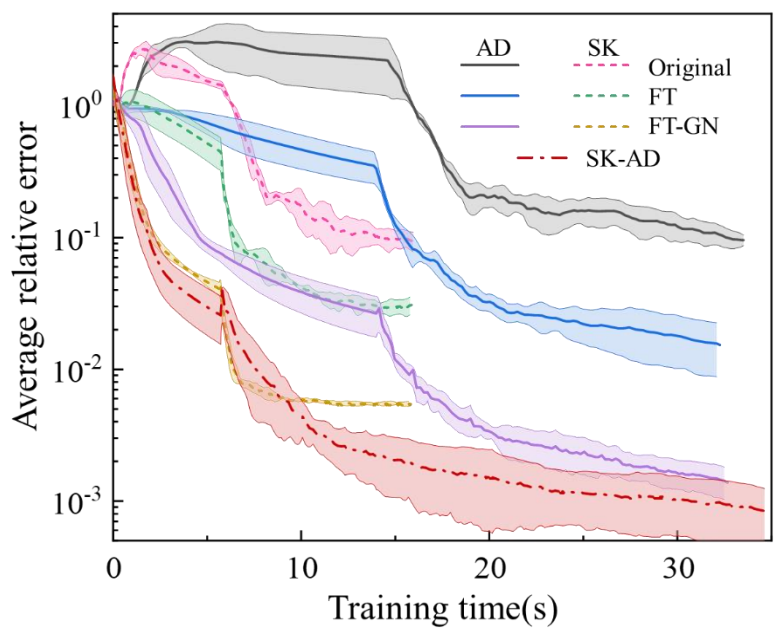

**Fig. 6.** The relationship between the average relative error of the approximate solution and training time for different forms of PINN in solving Poisson's equation.

Figure 6 illustrates the relationship between the average relative error of the approximate solution and training time for different variants of PINN in solving Poisson's equation. Solid lines depict the training results of AD-PINN, while dashed lines represent SK-PINN. "Original" denotes the baseline PINN method. "FT" refers to PINN using the random Fourier feature embedding technique from Section 2.2, and "FT-GN" indicates the combined use of random

Fourier features and adaptive loss balancing based on gradient norms as described in Section 2.3. Additionally, the red dot-dash line "SK-AD" represents SK derivatives applied with Adam for initial network training followed by AD with L-BFGS for further refinement.

The shaded areas in the Fig. 6 represent the standard deviation across three independent training runs, several advantages of SK-PINN over AD-PINN are evident. Firstly, under the same training settings (equal iterations for Adam and L-BFGS), SK-PINN requires less training time. Secondly, for an equivalent training duration, SK-PINN achieves a lower average relative error. Lastly, correction techniques for PINN are effective for SK-PINN, suggesting that SK derivatives can replace AD to expedite PINN training.

We suggest first rapidly training the network by using SK-PINN and then fine-tuning the solution by using AD-PINN. This approach acknowledges that while SK-PINN trains faster than AD-PINN, it may yield a coarser solution due to approximate derivatives, necessitating AD-PINN for refinement. Illustrated by the red dot-dash line, this "SK-AD" strategy facilitates rapid convergence to more accurate results.

## 5. Results and discussion

This section evaluated three types of PINNs. The first type, known as AD-PINNs, represents the original PINNs. The second type, termed mAD-PINNs, incorporates modification methods described in Section 2. The third type, referred to as SK-PINNs, utilizes Algorithm 1. The PINN architecture and training configurations used in our experimental studies are summarized in Table 2. For each test problem, the same network architecture and training settings were employed to compare the performance of AD-PINNs, mAD-PINNs, and SK-PINNs. The training cost for different models is shown in Table 2.

Table 2. PINN architecture and training configurations used in the experimental study and their training cost.

| Problem | 5.1 Inhomogeneous Poisson | | 5.2 Lid-driven cavity | 5.3 Flow around a cylinder | 5.4 Plate bending | 5.3 Elasticity imaging |
|---|---|---|---|---|---|---|
| | Square boundary | Complex boundary | | | | |
| Governing Eqs. | (39) | (39) | (44) | (44) | (45) | See [29] |
| PINN architecture | $(x,y)$-40-40-$(\hat{u})$ | $(x,y)$-40-40-$(\hat{u})$ | $(x,y)$-256-20-20-20-[20-20-20-$(\hat{u})$, 20-20-20-$(\hat{v})$, 20-20-20-$(\hat{p})$] | $(x,y)$-128-20-20-20-[20-20-20-$(\hat{u})$, 20-20-20-$(\hat{v})$, 20-20-20-$(\hat{p})$] | $(x,y)$-40-40-40-$(\hat{w})$ | $(x,y)$-256-20-20-20-[20-20-20-$(\widehat{\boldsymbol{S}})$, 20-20-20-$(\widehat{\boldsymbol{Lamé}})$] |
| $\Delta x$ (AD- & SK-) | $\Delta x = \Delta y = 0.01$ | $\Delta x = \Delta y = 0.02$ | $\Delta x = \Delta y = 0.004$ | - | $\Delta x = \Delta y = 0.02$ | - |
| $\sigma$ | 2 | 1 | 2 | 5 | 1 | 10 |
| Number of neighborhood points | 9 | 9 | 21 | 21 | 45 | 13 |
| Training sample (number of collocation points) | 40401+400 | 17835+419 | 63001+1592 | 23271+1160+251+251 | 10201+400 | 14200+284+200 |
| Max. training iteration | 1000+1000 | 1000+1000 | 20000+40000 | 20000+40000 | 10000+10000 | 5000+5000 |
| Total training cost (AD-/mAD-/SK-PINNs) | 1.6/1.6/0.33 mins | 0.84/0.81/0.21 mins | 57/67/1.2 h | 4.3/4.2/0.47 h | 52/53/2.7 mins | 10.3/9.9/2.8 mins |

1. For the PINN architecture, the numbers between input and output represent the number of nodes in each hidden layer. For example, $(x,y)$-40-40-$(\hat{u})$ indicates two inputs, $x$ and $y$ followed by 2 hidden layers with 40 nodes each, and a single output, $\hat{u}$.
2. We incorporate the random Fourier feature embedding into the first hidden layer of PINN and initialize its weights by sampling from a normal distribution $\mathcal{N}(0, \sigma^2), \sigma \in [1,10]$. These weights are kept constant during training.
3. Training samples contain different types of collocation points. For instance, the lid-driven cavity problem includes 63001 interior and 1592 boundary collocation points.
4. Max. training iteration refers to the respective training iterations for Adam and L-BFGS optimizers.

## 5.1. Poisson's equation

In this section, we present the results of the proposed SK-PINNs for solving the inhomogeneous Poisson's equation. This equation describes the governing PDE for the case of steady-state heat conduction with a source function, mathematically expressed as

$$\nabla^2 u(x,y) = f(x,y), \quad (x,y) \in \Omega. \tag{39}$$

Eq. (39) subject to Dirichlet boundary conditions. The network architecture is [2]+[40]+[40]+[1], indicating two input variables in the input layer, two intermediate hidden layers with forty neurons each, and one output variable. Training involves Adam for 1000 iterations, followed by L-BFGS for another 1000 iterations, with the loss weights updated every 100 iterations. Other network parameters are specified in Algorithm 1. If random Fourier feature embedding is used, the mapping of the second layer is modified according to (11) to maintain the network structure. It should be emphasized again that the difference between mAD-PINN and SK-PINN lies only in the calculation of derivatives.

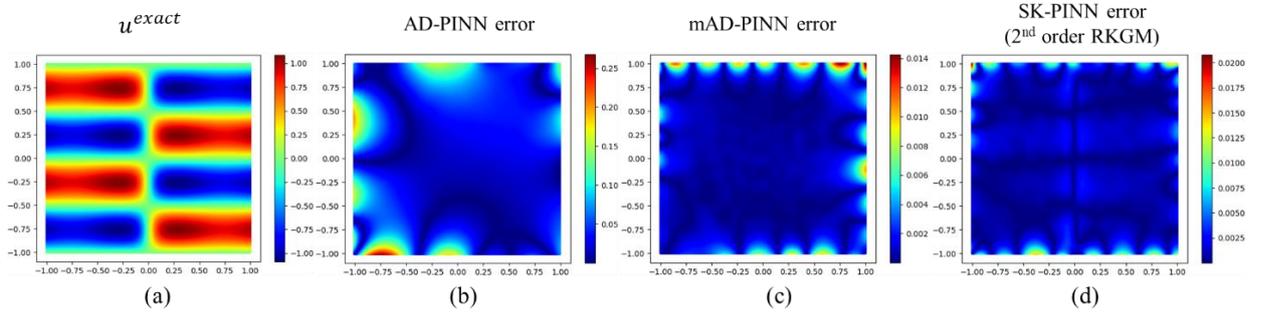

**Fig. 7.** The results of solving Poisson's equation with 40401 collocation points. (a) $u^{exact}$. (b) The absolute error of AD-PINN. (c) The absolute error of mAD-PINN. (d) The absolute error of SK-PINN.

**Table 3.** The errors and training cost of the three types of PINNs.

| Model | Number of neighborhood points | Number of collocation points | Rel. $L^2$ error (%) | Training cost (s) |
|---|---|---|---|---|
| AD-PINN | - | | 10.2/5.4/7.4 | 17.9/34.5/97.2 |
| mAD-PINN | - | | 0.24/0.29/0.32 | 17.5/32.9/94.6 |
| SK-PINN | 9 | 2601/10201/40401 | 2.8/0.71/0.23 | 9.2/11.0/20.1 |
| | 13 | | 1.9/0.86/0.36 | 9.2/11.2/21.2 |
| | 21 | | 1.8/0.57/0.35 | 9.2/11.6/22.4 |

**Example 5.1.1.** We solve Poisson's equation with the following exact solution, which exhibits a steep change along the $x$-direction and a sinusoidal behavior in the $y$-direction:

$$u^{exact}(x,y) = (0.1\sin(2\pi x) + \tanh(10x)) \times \sin(2\pi y), \Omega = [-1,1] \times [-1,1], \quad (40)$$

where the source function is obtained by substituting the exact solution into Eq. (39). Fig. 7 displays the results obtained by training three networks using 40,401 collocation points. It provides a color map of $u^{exact}$ and the absolute error color maps for each of the three models. Notably, the boundary error of AD-PINN is significantly larger than that of mAD-PINN and SK-PINN, while the errors of mAD-PINN and SK-PINN are of comparable magnitude.

To further analyze the performance of SK-PINNs, we conducted a detailed comparison of the training cost among the three models. Specifically, three scenarios were considered for SK-PINN, each with a different number of neighborhood points. The results are presented in Table 3, where the relative $L^2$ error is calculated using the following formula

$$e_{l2} = \sqrt{\frac{\sum (\hat{u} - u^{exact})^2}{\sum u^{exact\,2}}}. \quad (41)$$

The results demonstrate that, under identical configurations, SK-PINN is several times faster than AD-PINN and mAD-PINN, while achieving the same level of accuracy as mAD-PINN. As the number of collocation points increases, the relative $L^2$ errors of SK-PINN decrease across all three scenarios, indicating its superior convergence properties.

**Example 5.1.2.** To demonstrate the capability of SK-PINN in solving PDEs with complex geometric boundaries, we consider solving Poisson's equation with a curved boundary. The boundary of this problem is represented as follows

$$r = 1.5 + 0.14\sin(4\theta) + 0.12\cos(6\theta) + 0.09\cos(5\theta). \quad (42)$$

Here, $\theta \in [0, 2\pi)$, and the boundary points are obtained as $(x,y) = (0.02 + r\cos(\theta), r\sin(\theta))$. The exact solution is given by

$$u^{exact}(x,y) = e^x + e^y. \quad (43)$$

Figure 8 displays the color map of the exact solution and the absolute error color maps for the three models. It indicates that SK-PINN can effectively handle problems with complex geometric boundaries. Additionally, Table 3 provides detailed data on the relative $L^2$ errors and training time of the three models under different scenarios. With 17835 collocation points, the relative $L^2$ errors of all three models are less than 0.01%.

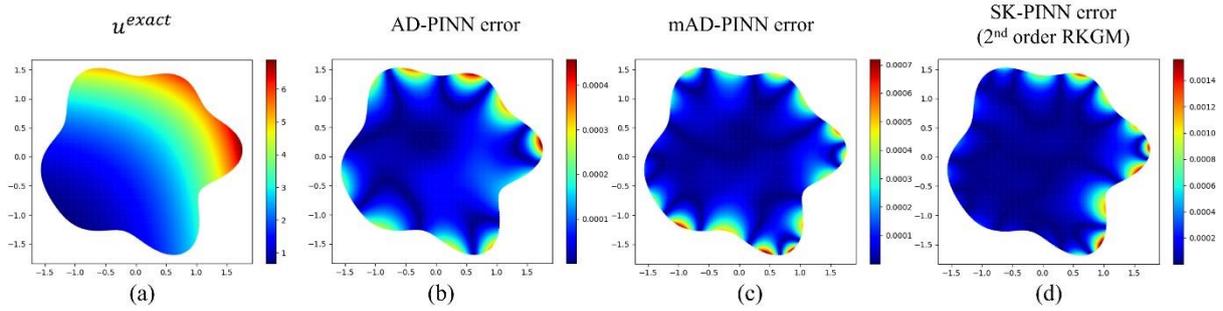

**Fig. 8.** The results of solving Poisson's equation with 17835 collocation points. (a) $u^{exact}$. (b) The absolute error of AD-PINN. (c) The absolute error of mAD-PINN. (d) The absolute error of SK-PINN.

**Table 4.** The errors and training cost of the three types of PINNs.

| Model | Number of neighborhood points | Number of collocation points | Rel. $L^2$ error (1e-2%) | Training cost (s) |
|---|---|---|---|---|
| AD-PINN | - | 709/2850/17835 | 0.20/0.35/0.43 | 12.5/18.7/49.8 |
| mAD-PINN | - | | 0.26/0.31/0.34 | 12.7/18.4/49.0 |
| SK-PINN | 9 | | 13.4/4.7/0.63 | 8.1/8.8/12.6 |
| | 13 | | 16.0/3.2/0.51 | 8.2/8.7/13.3 |
| | 21 | | 15.8/4.0/0.69 | 8.1/9.0/13.8 |

## 5.2. Lid-driven cavity

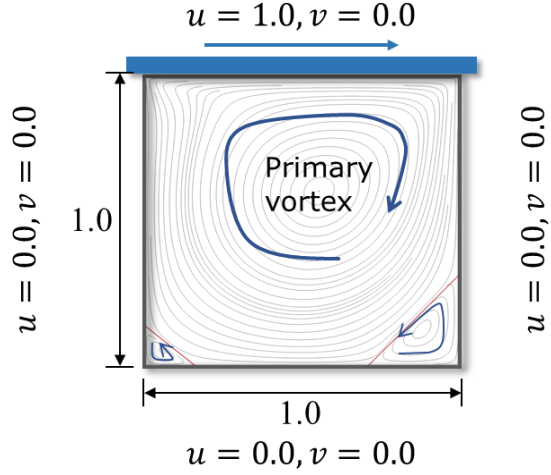

**Fig. 9.** Schematic of lid-driven cavity flow ($Re$=400).

The lid-driven cavity problem is widely used as a benchmark for numerical methods due to its complex physical characteristics. As shown in Fig. 9, this problem involves a unit square cavity where the lid at the top wall moves with a velocity of $u = 1$. The other walls are no-slip boundaries with $u = 0$ and $v = 0$. When the Reynolds number ($Re$) is less than 1000, two vortices appear in the lower right and left corners. The governing equations are the steady-state, two-dimensional, incompressible Navier-Stokes equations:

$$\begin{cases} \boldsymbol{u} \cdot \nabla \boldsymbol{u} + \nabla p - \dfrac{1}{Re} \Delta \boldsymbol{u} = \boldsymbol{0}, \\ \nabla \cdot \boldsymbol{u} = 0. \end{cases} \quad (44)$$

In these equations, the primary variables $\boldsymbol{u} = (u, v)$ and $p$ represent the velocity and pressure, respectively. $Re$ denotes the Reynolds number, the ratio of inertial forces to viscous forces. To avoid strong singularities at the boundary corners, we used the boundary implementation method described in [70].

Three PINNs were trained to solve the cavity flow at $Re = 400$. The detailed parameters for the three PINNs are listed in Table 2. Fig. 10a shows the true solution, while Figs. 10b-d show the absolute error distributions for AD-PINN, mAD-PINN, and SK-PINN compared to the true values. Table 5 presents the training time and relative $L^2$ errors for the different PINNs.

Both SK-PINN and mAD-PINN outperform AD-PINN, with mAD-PINN achieving the smallest error. SK-PINN shows larger errors at the upper left and right corners due to rapid velocity changes near these corners, causing derivative errors in the SK technique. These errors lead to inaccuracies elsewhere in the domain.

Although the SK derivative introduces errors, it shortens the backpropagation chain of the neural network, consuming less GPU memory. When using an RTX 3070 Ti GPU with 8 GB of memory, both AD-PINN and mAD-PINN exceeded the memory limit. SK-PINN's training speed was 55.8 times faster than mAD-PINN. The SK-PINN is also 4.75 times faster than the mAD-PINN using a 40GB NVIDIA Tesla A100-PCIe 4.0 GPU.

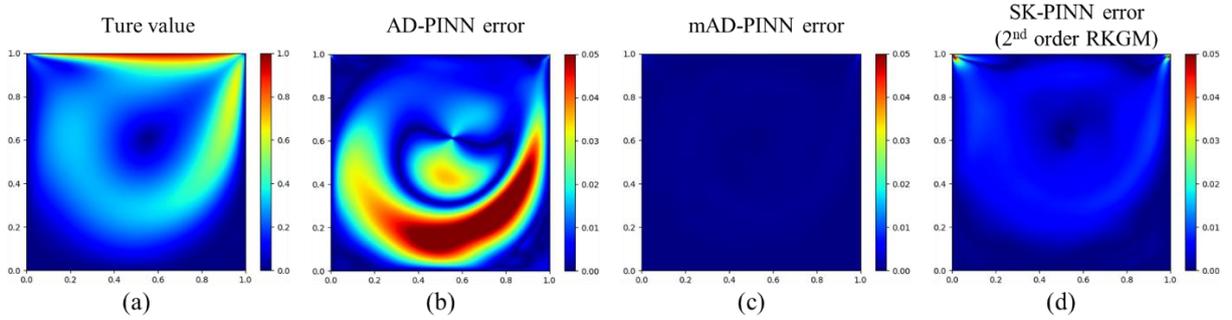

**Fig. 10.** The results of solving the Lid-driven cavity with 63001 collocation points. (a) True solution. (b) The absolute error of AD-PINN. (c) The absolute error of mAD-PINN. (d) The absolute error of SK-PINN.

**Table 5.** The training costs on different GPUs and the relative $L^2$ errors of different PINNs.

| Model | Number of neighborhood points | Number of collocation points | Rel. $L^2$ error (%) | Training cost (h) | |
|---|---|---|---|---|---|
| | | | | RTX 3070 Ti | Tesla A100 |
| AD-PINN | - | 63001 | 8.03 | 57 | 1.7 |
| mAD-PINN | - | | 0.21 | 67 | 1.9 |
| SK-PINN | 21 | | 1.88 | 1.2 | 0.4 |
| Coupled-PINN | 21 | | 0.46 | 1.8 | 1.1 |

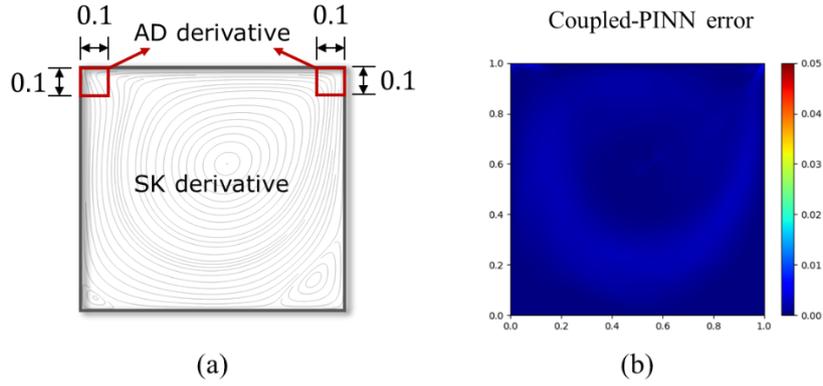

(a)     (b)

**Fig. 11.** (a) Illustrations of AD and SK derivatives in different regions. (b) The absolute error of Coupled-PINN compared to the true values.

To combine the advantages of SK-PINN and mAD-PINN, AD derivatives are used near the upper left and right corners. As shown in Fig. 11a, the region size is 0.1×0.1. While SK derivatives are used in other areas. This method is referred to as Coupled-PINN. The results for Coupled-PINN are shown in Fig. 11b. The training errors and time are listed in Table 5. Coupled-PINN has slightly lower accuracy than mAD-PINN, but its training time is significantly faster on different GPUs. Thus, the SK derivative offers a new option for training PINNs, allowing users to choose the appropriate derivative calculation method based on the problem characteristics.

5.3. Flow around a circular cylinder

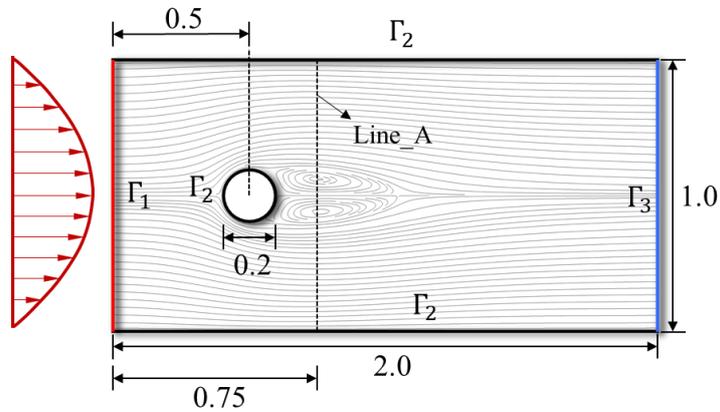

**Fig. 12.** Schematic of flow around a circular cylinder.

The flow around a cylinder is a frequently studied case in the literatures [15, 44, 61], governed by the Navier-Stokes equations. This flow situation exhibits a variety of patterns, such as vortex shedding and flow transition, which are highly dependent on the Reynolds number ($Re$). As the Reynolds number increases, more vortices are generated. In this case, the Reynolds number is set to $Re = 300$. The geometric parameters of the computational domain are detailed in Fig. 12. Within this domain, the governing PDEs are given by Eq. (44). The boundaries of the domain are categorized into three distinct types: $\Gamma_1$, $\Gamma_2$, and $\Gamma_3$. $\Gamma_1$ is the inlet boundary, where the velocity profile follows a parabolic distribution of $4y(1-y)$. $\Gamma_2$ is a no-slip boundary. $\Gamma_3$ is a Neumann boundary with $\frac{1}{Re}\frac{\partial u}{\partial x} = p$. When the flow reaches a steady state, two vortices appear behind the cylinder, which the streamline distribution is shown Fig. 12.

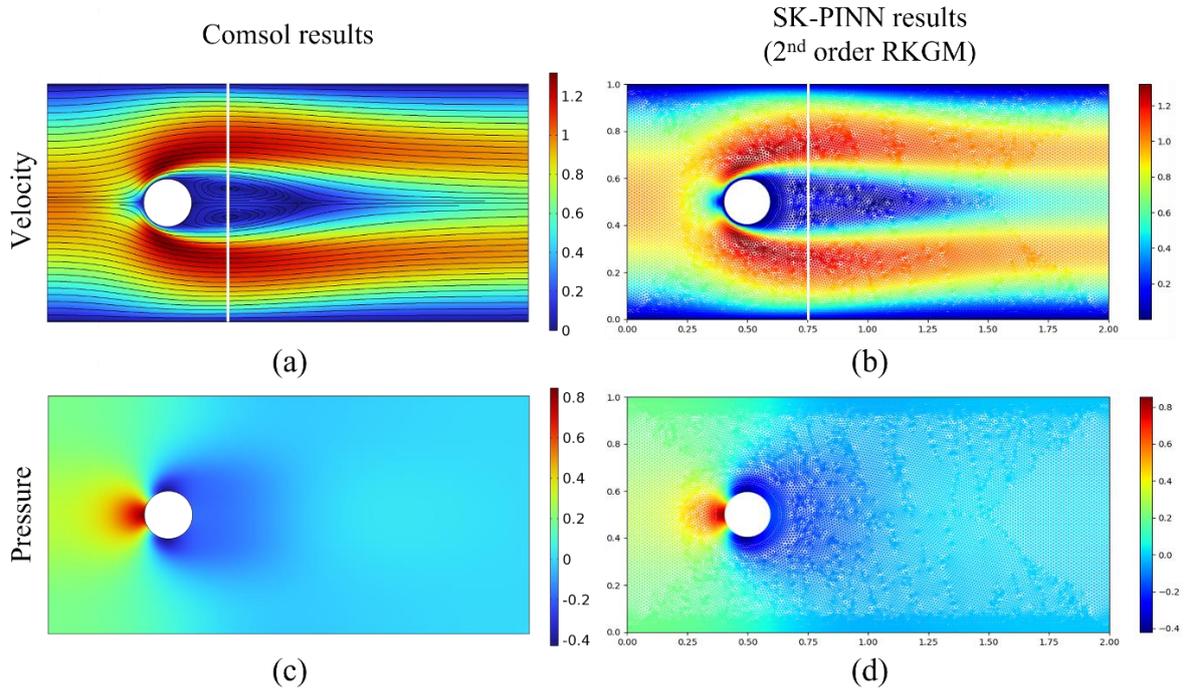

**Fig. 13.** Contour plot of velocity magnitude and pressure distribution.

The detailed parameter settings for this case are presented in Table 2. Fig. 13 compares the results of COMSOL Multiphysics 6.0, using a very fine mesh, with SK-PINN using non-uniform collocation points. Figs. 13a and 13b show the color maps of velocity magnitude, while Figs. 13c and 13d depict the color maps of pressure. To further quantitatively compare the accuracy differences between AD-PINN, SK-PINN, and mAD-PINN, we analyze Line_A at $x = 0.75$, which passes through the vortex region. Fig. 14 shows the velocity components along Line_A. Three PINNs produce vortices, with AD-PINN being less accurate than SK-PINN and mAD-PINN. mAD-PINN demonstrates the highest accuracy but requires 8.9 times the training time of SK-PINN. The training time for the three PINNs are listed in Table 2. Fig. 15 presents the relationship between the pressure around the cylinder and the arc length of the cylinder, leading to similar conclusions as mentioned above. In this case, SK-PINN performs well with non-uniform collocation points, verifying the reliability of the proposed algorithm.

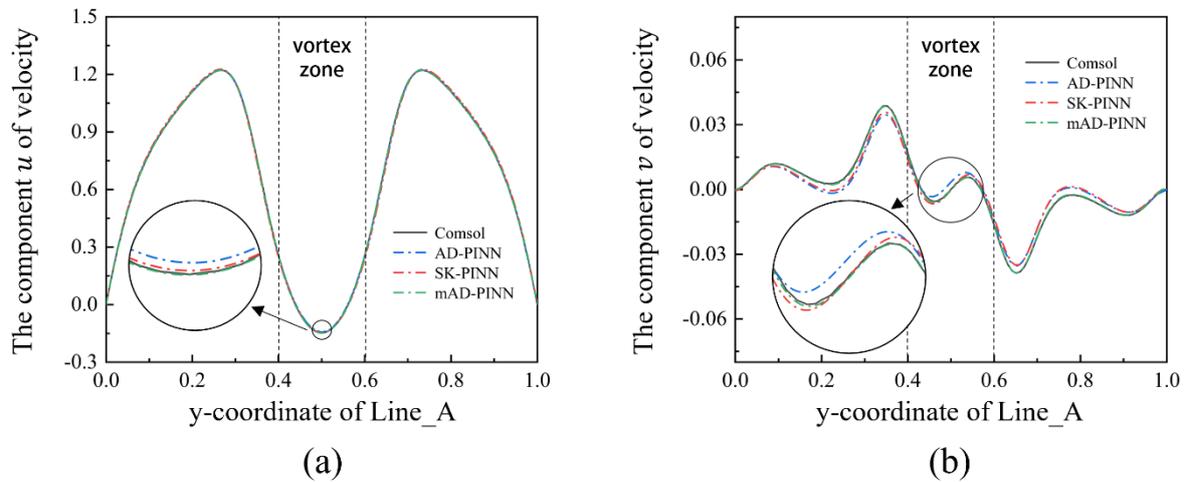

**Fig. 14.** Velocity components along the Line_A

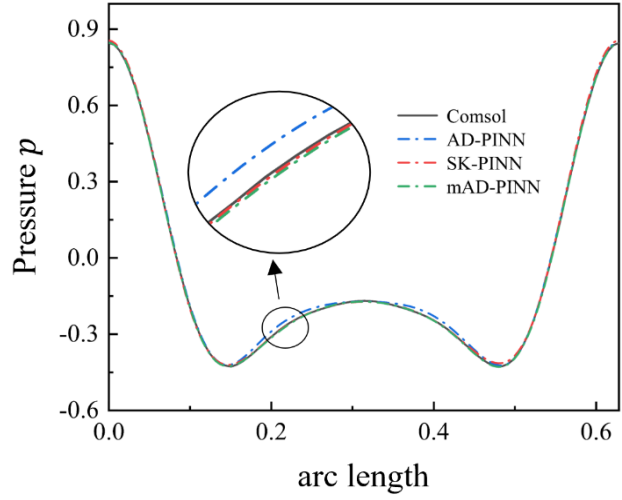

**Fig. 15.** Pressure Distribution around the Cylinder

5.4. Out-of-plane deflection of a square plate

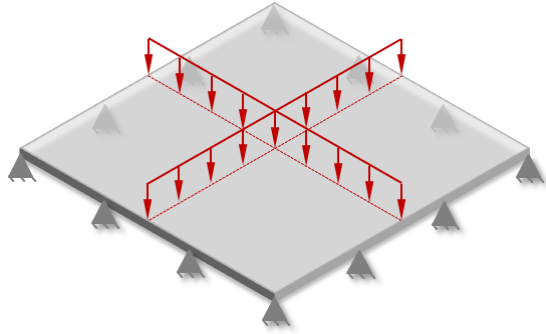

**Fig. 16.** Loading case of plate deflection under out-of-plane pressure

In this case, the deflection loading of a square plate under uniform out-of-plane pressure is considered. As shown in Fig. 16, a uniformly distributed transverse pressure is applied to the square plate, with all four edges clamped. The governing PDE is

$$D\nabla^4 w = q \tag{45}$$

Here, $D$ is the bending stiffness of the thin plate. It depends on the plate thickness, Young's modulus, and Poisson's ratio of the material. $D$ is set to 1 and $q$ is set to 10. The dimensions of the plate are $2 \times 2$. The boundary conditions are

$$w = 0, \frac{\partial w}{\partial n} = 0, (\text{at } x = \pm 1 \text{ or } y = \pm 1). \tag{46}$$

Finite element simulations were conducted using COMSOL Multiphysics 6.0 with an extremely fine element size. The training settings for SK-PINN are detailed in Table 2. The predicted deflection field is compared with the finite element simulation in Fig. 17. Fig. 17c shows the quantitative comparison along the centerline. The results demonstrate that SK-PINN accurately predicts the out-of-plane displacement. Notably, mAD-PINN requires step-by-step differentiation to solve this fourth-order differential equation, whereas 5th-order RKGM achieves all derivatives in a single step. Consequently, SK-PINN is 19.6 times faster than mAD-PINN for this problem. This case highlights the superior performance of SK-PINN in solving higher-order PDEs.

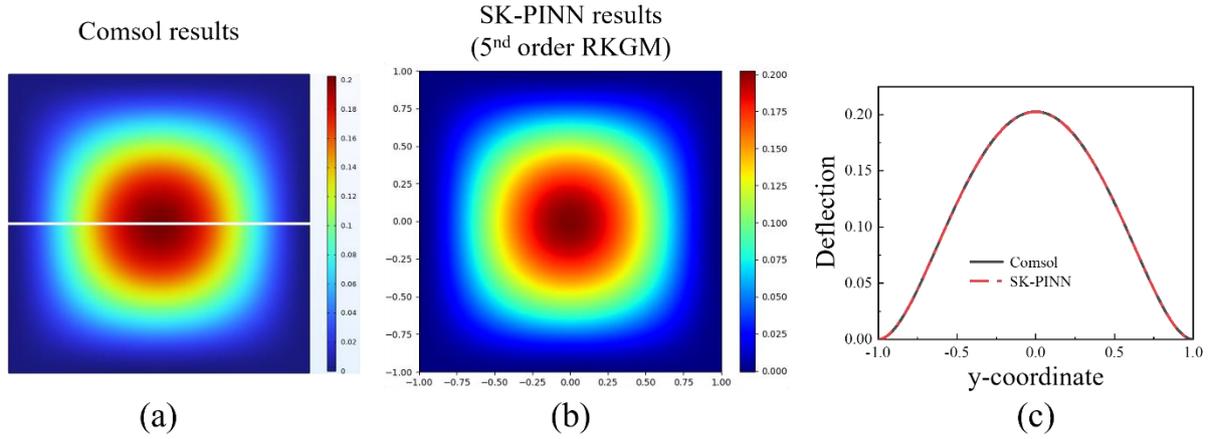

**Fig. 17.** The predicted deflection field is compared with the finite element simulation

## 5.5. Inverse identification of material parameters

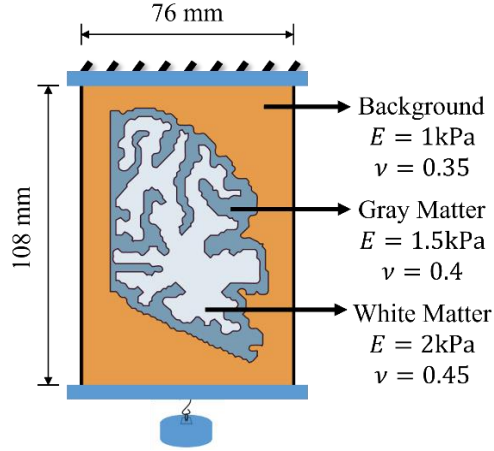

**Fig. 18.** Specifications of a brain slice with white matter and gray matter distinction, with varying $E$ and $\nu$ assignments for the three regions.

Elasticity imaging [29, 30] is a technique that utilizes deformation and force measurements under various loading conditions to reveal the spatial distribution of tissue mechanical properties. However, most existing methods only approximate a single material parameter, assuming a uniform distribution. In this case, SK-PINN was adopted to solve the linear elasticity problem. Strain data, normal stress boundary conditions, and physical equations are utilized for implementation, as detailed in [29]. SK-PINN accurately captures the spatial distribution of the elastic modulus ($E$) and Poisson's ratio ($\nu$), as well as tissue interfaces, using publicly available data from [29].

As shown in Fig. 18, a rectangular hydrogel sample undergoes uniaxial loading, which includes a human brain slice with distinct gray and white matter regions and complex geometric features. The distribution of different materials and the overall dimensions of the sample are provided in Fig. 18. Fig. 19 presents the results of SK-PINN in identifying the elastic modulus and Poisson's ratio from strain field data. The relative $L^2$ errors between the SK-PINN results and the true values for the elastic modulus and Poisson's ratio are 3.7% and 1.2%, respectively. Additionally, the training cost for the three PINNs are listed in Table 2. In this case, SK-PINN once again trains faster than the other two PINNs.

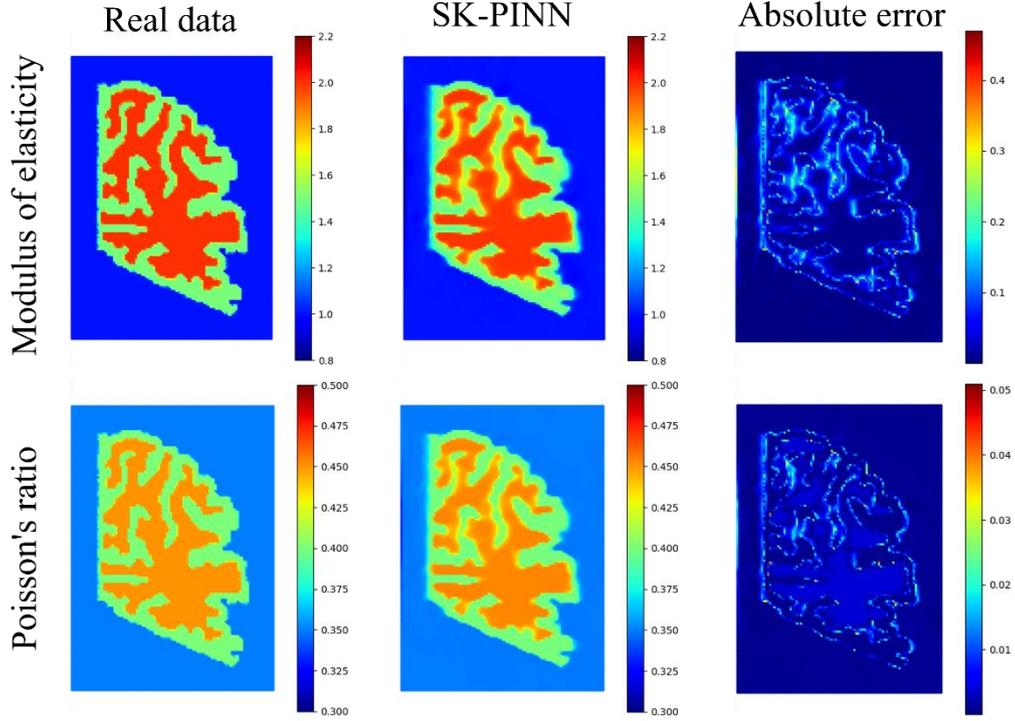

**Fig. 19.** Comparison of SK-PINN predicted material parameters and true material distribution

## 6. Conclusions

This paper presents the smoothing kernel-based physics-informed neural network (SK-PINN) and its application in computational mechanics. The combination of smoothing kernel functions and traditional physics-informed neural networks eliminates the computational bottleneck of automatic differentiation in conventional PINNs. Some main ideas and prominent features of the proposed approach are shown as follows:

- SK-PINN retains the advantages of meshless particle methods, allowing irregularly distributed collocation points. The strict mathematical convergence of derivative computation formulas allows to choose the order of derivative approximation freely according to problems.

- SK-PINN is efficient in large-scale collocation point problems, despite derivative approximation errors. Thus, a free combination of SK and AD derivatives accelerates

PINN training.

- Compared with traditional PINNs, SK-PINN has a flexible overall network architecture. For instance, non-differentiable neural network structures, such as convolutional neural networks, can also be combined with the proposed framework.

In future research on SK-PINN, here are several recommendations. First, causal learning strategies are introduced into SK-PINN to solve three-dimensional transient problems. Second, to avoid the spectral bias of fully-connected neural networks, SK-PINN should be extended to other network architectures, such as KAN networks. Lastly, weak-form or energy-form SK-PINNs could also be explored.

**Data availability statement**

The data that support the findings of this study are available from the corresponding author upon reasonable request.

**Acknowledgment**

The supports from the National Natural Science Foundation of China (12372192, 11972108, 12072061, 12072062), Key Research and Development Project of Liaoning Province (2020JH2/10500003) and Fundamental Research Funds for the Central Universities are gratefully acknowledged. The authors acknowledge the Supercomputing Center of Dalian University of Technology for providing computing resources.

[21] Y. Wang, J. Sun, J. Bai, C. Anitescu, M.S. Eshaghi, X. Zhuang, T. Rabczuk, Y. Liu, Kolmogorov Arnold Informed neural network: A physics-informed deep learning framework for solving PDEs based on Kolmogorov Arnold Networks, 2024, arXiv:2406.11045.

[22] K. Shukla, J.D. Toscano, Z. Wang, Z. Zou, G.E. Karniadakis, A comprehensive and FAIR comparison between MLP and KAN representations for differential equations and operator networks, Comput. Meth. Appl. Mech. Eng., 431 (2024) 117290.

[23] X. Jin, S. Cai, H. Li, G.E. Karniadakis, NSFnets (Navier-Stokes flow nets): Physics-informed neural networks for the incompressible Navier-Stokes equations, J. Comput. Phys., 426 (2021) 109951.

[24] R. Qiu, R. Huang, Y. Xiao, J. Wang, Z. Zhang, J. Yue, Z. Zeng, Y. Wang, Physics-informed neural networks for phase-field method in two-phase flow, Phys. Fluids, 34 (2022) 052109.

[25] Z. Wu, H. Zhang, H. Ye, H. Zhang, Y. Zheng, X. Guo, PINN enhanced extended multiscale finite element method for fast mechanical analysis of heterogeneous materials, Acta Mech., (2024) 1-19.

[26] Y. Wang, J. Sun, W. Li, Z. Lu, Y. Liu, CENN: Conservative energy method based on neural networks with subdomains for solving variational problems involving heterogeneous and complex geometries, Comput. Meth. Appl. Mech. Eng., 400 (2022) 115491.

[27] P. Zhang, Y. Hu, Y. Jin, S. Deng, X. Wu, J. Chen, A Maxwell's equations based deep learning method for time domain electromagnetic simulations, IEEE J. Multiscale Multiphysics Comput. Technol., 6 (2021) 35-40.

[28] M. Raissi, A. Yazdani, G.E. Karniadakis, Hidden fluid mechanics: Learning velocity and pressure fields from flow visualizations, Science, 367 (2020) 1026-1030.

[29] A. Kamali, M. Sarabian, K. Laksari, Elasticity imaging using physics-informed neural networks: Spatial discovery of elastic modulus and Poisson's ratio, Acta Biomater., 155 (2023) 400-409.

[30] A. Kamali, K. Laksari, Physics-informed UNets for discovering hidden elasticity in heterogeneous materials, J. Mech. Behav. Biomed. Mater., 150 (2024) 106228.

[31] Z. Fang, J. Zhan, Deep physical informed neural networks for metamaterial design, IEEE Access, 8 (2019) 24506-24513.

[32] S. Wang, Y. Teng, P. Perdikaris, Understanding and mitigating gradient flow pathologies in physics-informed neural networks, SIAM J. Sci. Comput., 43 (2021) A3055-A3081.

[33] S. Wang, X. Yu, P. Perdikaris, When and why PINNs fail to train: A neural tangent kernel perspective, J. Comput. Phys., 449 (2022) 110768.

[34] L.D. McClenny, U.M. Braga-Neto, Self-adaptive physics-informed neural networks, J. Comput. Phys., 474 (2023) 111722.

[35] M.A. Nabian, R.J. Gladstone, H. Meidani, Efficient training of physics-informed neural networks via importance sampling, Comput.-Aided Civil Infrastruct. Eng., 36 (2021) 962-